\documentclass[10pt,twocolumn,twoside]{IEEEtran}
\usepackage[all]{xy}
\usepackage{color}
\usepackage{algorithm}
\usepackage{algpseudocode}
\usepackage{subfigure}
\usepackage{bm}
\usepackage{graphicx}
\usepackage{color}
\usepackage{wrapfig}
\usepackage{epsf}
\usepackage{mathptm}
\usepackage{times}
\usepackage{url}
\usepackage{amssymb, amsmath, mathtools, amsthm}
\usepackage{nomencl}
\usepackage{comment}
\usepackage{enumitem}
\usepackage{hyperref}
\makenomenclature

\definecolor{RED}{rgb}{0.6,0.,0.}
\definecolor{BLUE}{rgb}{0.,0.,0.6}
\definecolor{GREEN}{rgb}{0.,0.6,0.}
\definecolor{MALINA}{rgb}{0.6,0.,0.6}
\definecolor{YELLOW}{rgb}{0.8,0.8,0}

\usepackage{subfigure,tikz}
\usetikzlibrary{backgrounds,automata}
\usepackage{tkz-tab}
\usepackage{pgfplots}
\pgfplotsset{compat=newest}
\pgfplotsset{every tick label/.append style={font=\footnotesize}}
\usepackage{csvsimple}
\usepackage{cancel}
\usepackage{amssymb}

\usepackage[hyphenbreaks]{breakurl}
\graphicspath{{./figures/}{./}}

\def\C{{\mathbb C}}

\def\H{{\mathsf{H}}}

\begin{document}
\title{Learning from power system data stream:\\ phasor-detective approach} 
\author{\IEEEauthorblockN{Mauro Escobar, 
Daniel Bienstock* \& Michael~Chertkov\dag\\}
\IEEEauthorblockA{*Columbia University, NY, USA\\
\dag Los Alamos National Laboratory, NM, USA \&  Skoltech, Moscow, Russia}
\thanks{Mauro Escobar, 
Daniel Bienstock are with Columbia University, USA. Email: \{me2533, dano\}@columbia.edu}
\thanks{Michael Chertkov is with Los Alamos National Laboratory, Los Alamos, NM and Skolkovo Institute of Science and Technology, Russia. Email: chertkov@lanl.gov}
}
\maketitle
\begin{abstract}
Assuming access to synchronized stream of Phasor Measurement Unit (PMU) data over a significant portion of a power system interconnect, say controlled by an Independent System Operator (ISO), what can you extract about past, current and future state of the system? We have focused on answering these practical questions pragmatically --- empowered with nothing but standard tools of data analysis, such as PCA, filtering and cross-correlation analysis. Quite surprisingly we have found that even during quiet ``no significant events" periods this standard set of statistical tools allows the ``phasor-detective" to extract from the data important hidden anomalies,  such as problematic control loops at loads and wind farms, and mildly malfunctioning assets, such as transformers and generators. We also discuss and sketch future challenges a mature phasor-detective can possibly tackle by adding machine learning and physics modeling sophistication to the basic approach.  
\end{abstract}

\section{Introduction}%

Data-driven techniques in power systems have at least fifty years of history, starting with static state estimations developed by Schweppe and co-authors \cite{70SW,70SR,70Sch}, then transitioning to dynamic state estimation analysis and applications, see e.g.  \cite{1990Hauer,2007Mani,12Tru} and references therein, and most recently discussed under the umbrella of ``big data'' as the most significant enabler of power system operations, security  and resiliency in the future \cite{13KXG,18AZ}. (See also related discussion in the description of the US Department of Energy new funding opportunity to ``explore the use of big data, artificial intelligence, and machine learning technology to leverage the power of grid sensors" \cite{DOE_call}.) Many specific questions and approaches,  including but not limited to modes of oscillations analysis of stability and detection \cite{12Tru,NERC_reliability,PNNL_oscil,WECC-JSI}, dimension reduction for faster processing and analysis \cite{12DKM}, early event detection \cite{14XCK}, missing data recovery \cite{16_RPI_missing_data}, identification of cyber attacks \cite{16_RPI_cyber} and real time (online) event detection \cite{18LWC} are among the most recent advances. 

On the methodology side data-driven methods developed in other engineering disciplines have been adopted, modified and used for many (e.g. aforementioned) power system applications. Principal component analysis 
\cite{12DKM}-\cite{18LWC}, auto-correlation analysis of memory effects \cite{18RL}, and linear model driven spectral analysis of the dynamic state matrix \cite{2007Mani,12Tru,NERC_reliability,PNNL_oscil,WECC-JSI,2009Chang,2016Mani,2017LVSDC} are arguably the most popular data-driven techniques currently in use in the power system research.

Even though the sophistication level of the methods already used in power system applications is impressive, coherence and understanding of the potential of new generation of the big data methods, driven during the last decade largely through heavy investment of IT industry, is still lacking \cite{DOE_call}. We anticipate that many of the most modern methods, especially Deep Learning (DL) and related techniques linked to Machine Learning (ML) and  Artificial Intelligence (AI), revolutionary advances in data science and more generally theoretical engineering \cite{09Ben,10ARK,15Sch,15LBH,16GBC}, will impact the power-system operation-room reality in an even more significant ways. However, one problem with applications of the novel methods of DL and alike in sciences and engineering is that they are application agnostic/generic -- very effective for many business cases, but lacking ``explainability", i.e. intuitive physical/engineering explanations.  This significant handicap of the most advanced and recent ML \& AI methods slows down development of related applications in power systems.  Indeed,  power system practitioners would generally not consider as practical any new methods lacking ``power systems informed" explanations.

This manuscript takes a step towards closing the gap between the rich variety of methods already developed and utilized in power systems and yet to be unleashed power of the upcoming Big Data revolution. Specifically, we start walking towards exciting sophistication of DL slowly, from the well-established and intuitive trenches of practical system engineering. We develop in this manuscript a pragmatic ``phasor-detective" approach to analysis of the streaming Phasor Measurement Unit (PMU) data which allows to extract and interpret spatial and temporal correlations in a computationally light fashion and without making any constraining assumptions about origin of the correlations. 

\subsection{Our Contribution}

We analyze synchronized historical PMU data recorded  at $\approx 200$  most significant locations of a US Independent System Operator (ISO) over the course of two years. At each PMU location the data includes complex current and voltage recorded with a millisecond resolution. Given geo-spatial locations of the PMU, but no information about the grid characteristics and layout, we pose the following principal questions: What can we possibly reconstruct from the data stream about the system current ambient behavior? To answer the question we utilize available statistical tools. In relation to preliminary data processing we apply to the raw signal three filtering techniques: moving average, sliding time horizon and Fourier analysis pre-processing. This allows to provide robust identification of the ``quiet" periods and also prepare data for subsequent statistical analysis by means of Principal Component Analysis (PCA) and Auto-Correlation Analysis (ACA). We show that the two complementary tools, applied to the raw as well as to pre-processed signal, allow to separate scales and also provide compressed, thus easy to visualize, descriptors for online tracking of current state of the grid in a much broader way than what the current Energy Management System (EMS) actually uses. PCA provides a robust set of indicators which record slow/adiabatic changes on the scale of seconds to ten of seconds and slower. We have observed that only a very few principal modes are significant at any moment of time, even though these modes may be different for voltage amplitude, phase difference and frequency (the three main characteristics) we track. The results do not change when the PCA is applied to the filtered signal, consistently with the fact that PCA averages over time but do not catch different-time correlations. ACA is the tool used to analyze the latter, in particular identifying significant, persistent correlations, missed by PCA, at shorter time scales - subseconds-to-seconds.  Following ACA curves at different spatial locations we were able to identify nodes where correlations do not decay with time showing significant memory-effects. Remarkably,  these nodes with significant memory cluster geographically. We observe two areas in the grid which show especially strong sustainable temporal correlations. We then proceed with ACA analysis of the Fourier-filtered signal.  This helps us to identify and localize different harmonics. In particular, we observe (for a particular quiet period) emergence of significant oscillations in the 4-6Hz range at a small number of nodes. Interestingly, nodes with significant sustainable oscillations are either wind farms, big aggregated loads or mid-size generators. We conjecture that the sustainable oscillations are indicators of malfunction at these critical elements of the grid. We also observe that sustainable oscillations, seen clearly through emergence of a residue in the ACA analysis of the raw signal, disappear when applied to the Fourier-filtered signal (cutting off the 4-6Hz oscillations). Finally, analyzing spatial cross-correlations (of the residue) we were able to identify group of nodes with significant inter-dependency.

Material in the manuscript is organized as follows. Logic and main steps of the ``phasor detective" approach are described in Section \ref{sec:logic}. Section \ref{sec:time_series} discusses the time series data we are working with and it also introduces averaging and filtering approaches we apply to the data. Covariance matrix and PCA analysis are discussed in Section \ref{sec:PCA}. Auto-correlation and cross-correlation approaches, extracting information of the data temporal correlations/memory,  are developed in Section \ref{sec:temp-correlations}. Section \ref{sec:conclusions} and Section \ref{sec:path-forward} are reserved, respectively, for conclusions and discussions of the path forward.

\section{Logic and Steps of the Detective Approach}
\label{sec:logic}

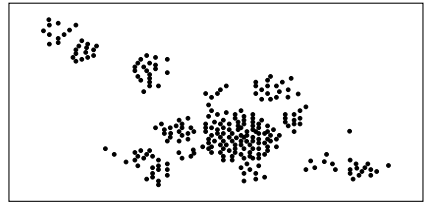
\begin{figure}
  \begin{center}
    \begin{tikzpicture}
    \begin{axis}[%
        ticks=none,
    	width=0.8\linewidth,
    	height=120pt,
        ]
        \addplot[only marks,mark size=0.7pt]%
          table[x=x,y=y,col sep=comma]{newcoords.csv};
    \end{axis}
    \end{tikzpicture}
  \caption{Geographical location of PMUs (anonymized coordinates). \label{fig:death_valey}}
  \end{center}
\end{figure}

In this work we report on data streams from over $200$ PMUs operated by an ISO and spanning a period over one year long. Fig.~\ref{fig:death_valey} displays a rendition of the locations of the PMUs using anonymized coordinates. As is standard, 
the data stream from each PMU includes (complex) current and voltage reported 30 times per second.
Using this data one can obtain real-time estimates of complex power at each location.   Working with a data set this large (on the order of 28 TB) presents some obvious challenges; additionally there are specific artifacts that can arise in the data.  For example, not all PMUs are always reporting, and occasionally some PMUs exhibit what appears to be errant behavior.  

Our work has centered on performing statistical analysis aimed at inferring ``structure" in the underlying transmission system as well as identifying complex behaviors, such as resonance and oscillations.  
In this manuscript we focus specifically on identification and characterization of ``quiet" periods. Oversimplifying (see related discussion below) we consider a period \textit{quiet} if fluctuations around the mean (e.g. characterized in terms of the standard deviations) are smaller than a reasonable pre-defined threshold.  This focus on the quiet/ambient periods is motivated by the following considerations:
\begin{itemize}
    \item The development of a strong understanding of quiet periods and (in particular) efficient online algorithms for recognition of such periods is a necessary step prior to studying less-quiet or even anomalous regimes, for otherwise we risk significant misinterpretation, i.e. errors in online detection of anomalies. 
    \item As will be seen in the following,  the quiet regimes display informative patterns and correlations, all (slowly) time-evolving.  Identifying such features is important with regards to
    \begin{itemize}
        \item Developing fast and reliable identification techniques.
        \item Uncovering hidden malfunction of assets thus providing significant contribution towards forecasting most probable (and destined to occur) failures. 
    \end{itemize}
    \item The richness of correlations observed in the quiet regime, in fact, suggests that separation of what is normal/quiet from what is anomalous/atypical will be challenging.  Even though we observe that quiet regimes dominate, relatively abrupt jumps of moderate size (i.e. jumps exceeding currently tracked standard deviation by factor of two or three) are rather frequent even though overall they account for a relatively short fraction of the stream.  As a result, it is rather difficult to find sufficiently long completely quiet periods in the available data.
    \item Clearly, understanding quiet vs volatile behavior will be helpful toward building predictive models for better optimization, control and planning. 
    \item ``Cyber" and ``cyber-physical" attacks on power systems are a venue where fast and effective learning of (changing) stochastics may prove useful in identifying attacks.  See \cite{2018attack}.
\end{itemize}
The methodology adopted in this manuscript to identify the quiet periods is explained in Section \ref{sec:quiet}.

{\bf Notation:} for $x\in\C$, $|x|$ denotes its magnitude and $\overline{x}$ denotes its complex conjugate. For a complex matrix $A$, $A^\H$ denotes its Hermitian transpose.

\section{Description and Averaging of the \\Time Series Data}
\label{sec:time_series}

The available data encompasses the period from January 1st 2013 to March 21st 2014. Each of the $N = 240$ PMUs records the following measurements 30 times per second: time of the measurement (GPS tagged), bus ID, voltage amplitude, voltage phase angle, current magnitude, current angle, frequency, and additional data not used in this study. Additionally, the 2-dimensional coordinates of the PMU locations is also available, together with their corresponding nominal voltages.  We note that PMUs \textit{report} 30 times/second, but they
\textit{sample} at a far higher rate and perform filtering (e.g., anti-aliasing) before reporting. 


We will denote a generic scalar or complex measurement (e.g. complex voltage) at PMU location $i$, at time $t$ by  $m_i(t)$. The parameter $t$ will be used to refer to the discrete time sequence with each temporal data point separated from preceding one by the same duration $\Delta$ ($1/30^{th}$ of a second in our case). 

Typical pre-processing steps in the statistical analysis of data (especially with the goal of analyzing correlations) involve modifications through de-trending, offsetting (subtraction) of moving average, and normalizations.  We will apply such techniques below.
Specific details are provided next. 

\subsection{Moving average and covariance}

The \emph{moving average} $\mu^{(m)}$ and \emph{moving variance} $\sigma^{(m)}$ of a parameter $m$ is computed at every bus $i$ in the following way:
\begin{align}
    \mu_i^{(m)}(t;\alpha) &= \alpha\cdot m_i(t) + (1-\alpha)\cdot \mu_i^{(m)}(t-1;\alpha),  \label{eq:movinga} \\
    \sigma_i^{(m)}(t;\alpha) 
    &= \alpha\cdot |m_i(t)-\mu_i^{(m)}(t-1;\alpha)|^2 \nonumber\\
    &\quad + (1-\alpha)\cdot \sigma_i^{(m)}(t-1;\alpha),
    \label{eq:movingv}
\end{align}
with some initial values, say $\mu_i^{(m)}(0;\alpha)=0$ and $\sigma_i^{(m)}(0;\alpha)=1$.
The parameter $\alpha\in(0,1)$ represents the degree of weighting decrease. Note that in \eqref{eq:movingv} we are using the moving average defined in  \eqref{eq:movinga}.  One can likewise define moving covariance parameters.

Finally, we introduce the zero mean and normalized zero mean data streams
\begin{eqnarray}
    && \bar{m}^{(m)}_i(t;\alpha) = m_i(t)-\mu_i^{(m)}(t-1;\alpha),\label{eq:mov_av}\\
    && \hat{m}^{(m)}_i(t;\alpha) = \frac{m_i(t)-\mu_i^{(m)}(t-1;\alpha)}{\sqrt{\sigma_i^{(m)}(t-1;\alpha)}},
    \label{eq:mov_av_norm}
\end{eqnarray}
obtained from the input stream by making use of the moving average and variance.  The zero mean parameters will help us identify quiet periods, as discussed next.

\subsection{Quiet Periods}
\label{sec:quiet}

Given a reference time $t$ and a length parameter $Q$  consider the $N\times Q$ matrix of normalized measurements
\begin{equation*}
    M(t;\alpha;Q) = [\ \hat{m}_i^{(m)}(\tau;\alpha)\ |\ \forall i,\ \tau\in\{t-Q+1,\ldots,t\}\ ],
\end{equation*}
corresponding to the last $Q$ measurements before time $t$ for all buses.
We define the period $(t-Q,t]$ as \emph{quiet} if the absolute value of all entries of the matrix $M(t;\alpha;Q)$ is below some preset threshold. The reason behind this definition is that sudden jumps in the data appear as large values in the normalized time series, whereas normalized values close to zero mean that the data is behaving in a steady way. 
Effectively, a quiet period is an interval of time where all sensor-reported data behave in a stationary way.
Moreover it is relatively cheap to compute $M(t;\alpha;Q)$ from the stream data, as we just have to keep track of the moving average and moving variance of each bus, and the matrix of the last $Q$ normalized measurements.

In our analysis, we used $\alpha=0.05$ as the weighting decrease constant of the moving average and moving variance. This value of $\alpha$ corresponds to the memory budget (in the moving average) of approximately one second, see Fig.~\ref{fig:averages}.  In our analyses we will consider quiet periods spanning $15$ minutes ($Q=27,000$).  Over a selection of five different days across the database, we compute we compute the matrix $M(t;\alpha;Q)$ and record its maximum absolute value when $t$ spans over the complete day. Just in few cases the maximum was below 10 units, we selected 2-3 intervals between these recorded cases.

Averaging over a fixed sliding window, described next for completeness, is an alternative to the moving window average discussed so far.

\subsection{Averaging over Sliding Time Horizon}
\label{subsec:Sliding}

\begin{figure*}
  \includegraphics[width=\textwidth]{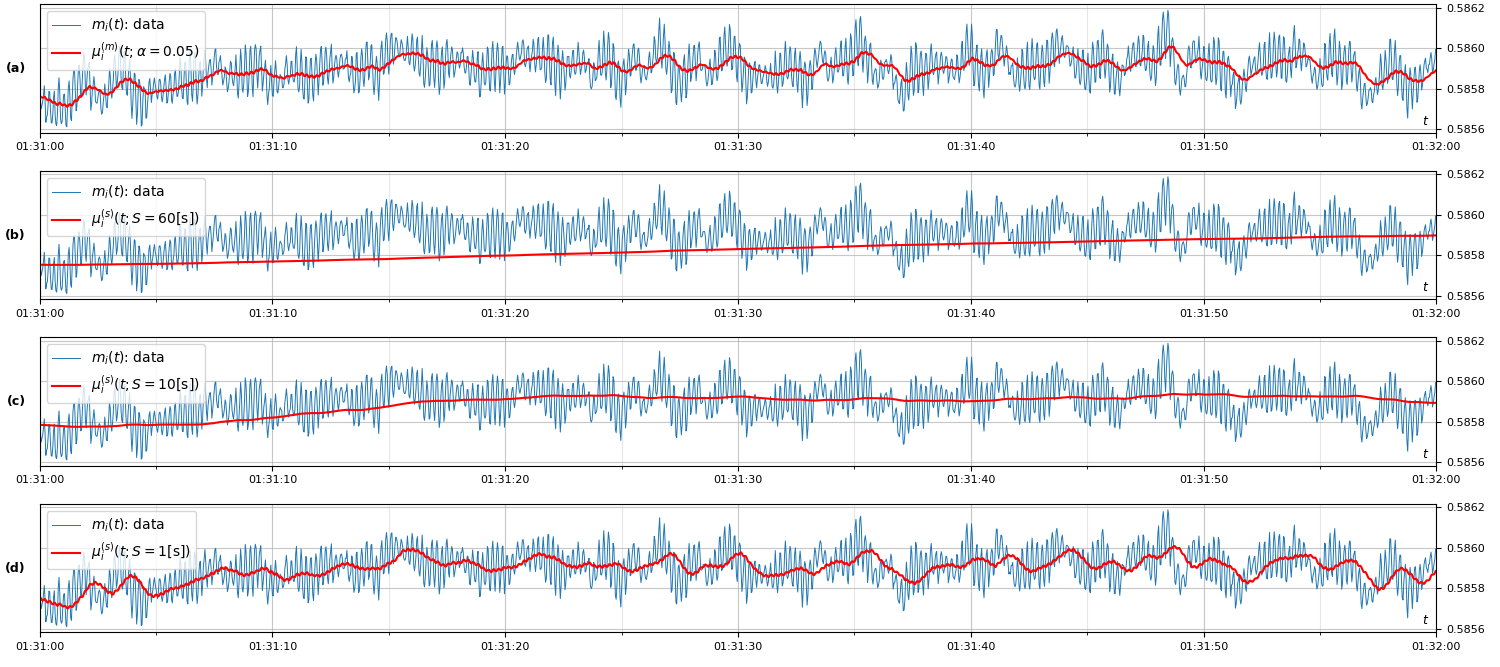}
  \caption{Raw data $m_i(t)$ during 1 minute of voltage magnitude at a particular bus and the comparison of the {\bf(a)} moving average $\mu_i^{(m)}$ with $\alpha=0.05$ and the window average $\mu_i^{(s)}$ for $S$ equal to {\bf(b)} 60, {\bf(c)} 10, and {\bf(d)} 1 seconds. 
  \label{fig:averages}}
\end{figure*}

We continue to work with the scalar or complex signal (voltage, phase or frequency) data stream $m_i(t)\ \forall i$, with $t\in\{0,\ldots,Q-1\}$, corresponding to the (not normalized) measurements of a selected period. Here, $t=0$ indicates the beginning of the period.
$N$ dimensional vectors of means and variances, averaged over the (last) sliding time horizon of duration $S\leq Q$ are defined as follows, $\forall t \mbox{ such that } S\leq t<Q$:
\begin{eqnarray}
&& \mu^{(s)}_i(t;S)= \frac{1}{S}\sum_{\tau=t-S+1}^t m_i(\tau), \label{eq:mu-s}\\
&& \sigma^{(s)}_i(t;S)= \frac{1}{S}\sum_{\tau=t-S+1}^t \left|m_i(\tau)-\mu_i^{(s)}(t;S)\right|^2,\label{eq:sigma-s}
\end{eqnarray}
then, the zero-mean data vector and the re-scaled zero-mean data vector are given by
\begin{eqnarray}
&& \bar{m}^{(s)}_i(t;S)=  m_i(t)-\mu_i^{(s)}(t-1;S),\label{eq:bar-m-s}\\
&& \hat{m}^{(s)}_i(t;S)=  \frac{m_i(t)-\mu_i^{(s)}(t-1;S)}{\sqrt{\sigma_i^{(s)}(t-1;S)}}.\label{eq:hat-m-s}
\end{eqnarray}
Numerically, we found that a reasonable choice for $S$, i.e. one consistent with $\alpha=0.05$ which allows to separate power electronics (milliseconds) and electro-mechanical (seconds) time scales, is the number of readings in 1 second, that is $S=30$. See Fig.~\ref{fig:averages}. 

\begin{figure*}
  \includegraphics[width=\textwidth]{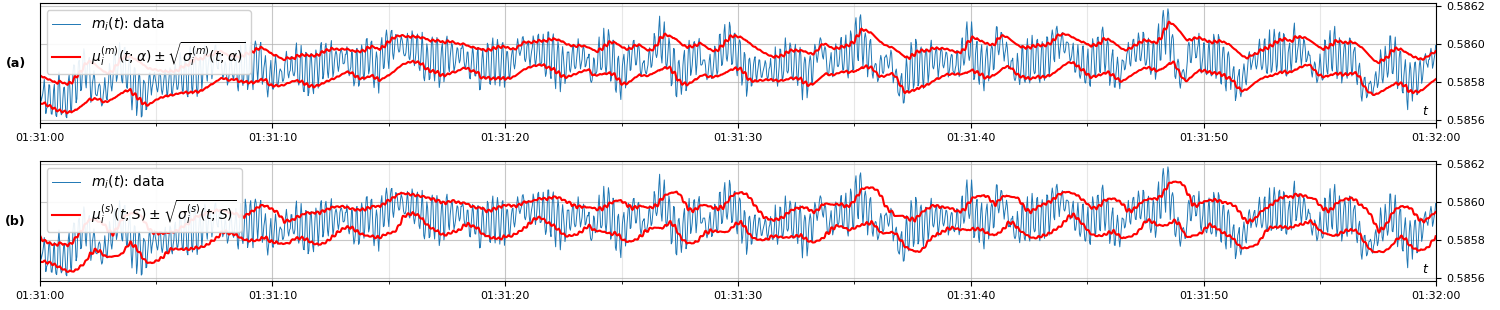}
  \caption{Raw data $m_i(t)$ during 1 minute of voltage magnitude at same bus than Fig.~\ref{fig:averages} and the comparison of {\bf(a)} $\mu_i^{(m)}\pm\sqrt{\sigma_i^{(m)}}$ with $\alpha=0.05$ and {\bf(b)} $\mu_i^{(s)}\pm\sqrt{\sigma_i^{(s)}}$ using $S=1$ second.
  \label{fig:std}}
\end{figure*}

Fig.~\ref{fig:std} shows a comparison between the behavior of $\sigma^{(m)}(t,\alpha=0.05)$ and $\sigma^{(s)}(t;S=30)$ --in the same time windows as Fig.~\ref{fig:averages}-- showing $\mu^{(\cdot)}\pm\sqrt{\sigma^{(\cdot)}}$.

\subsection{Fourier Filter}
\label{sec:FTfilter}

We have also applied Fourier filtering on the time series $m_i(\cdot)$. Let ${\cal F}[m_i(\cdot)]\in \C^Q$ be the discrete Fourier transform of $m_i(\cdot)$, where ${\cal F}[m_i(\cdot)]_k$ is the amplitude corresponding to frequency $\omega_k$, $k\in\{0,\ldots,Q-1\}$. 
Since the frequency of the readings is $30$Hz, we can represent the frequency domain as $Q$ equidistant points between $-15$ and $15$Hz.

We obtain a filtered time series by suppressing the high frequency components. 
Let ${\bf 1}_{[-\lambda,\lambda]}\in\{0,1\}^Q$ be the indicator function of the set of frequencies in the interval $[-\lambda,\lambda]$, that is
\begin{equation*}
    ({\bf 1}_{[-\lambda,\lambda]})_k = \begin{cases}
    1, &\mbox{if }-\lambda\le \omega_k\leq \lambda\\
    0, &\mbox{otherwise.}
    \end{cases}
\end{equation*}
The filtered series is ${\cal F}^{-1}\left[{\bf 1}_{[-\lambda,\lambda]}\odot{\cal F}[m_i(\cdot)]\right]$, where ${\cal F}^{-1}$ is the inverse Fourier transform and $\odot$ is the component-wise product. We will denote this new series by $f_i(t;\lambda)$, $t\in\{0,\ldots,Q-1\}$.  

\begin{figure*}
  \includegraphics[width=\textwidth]{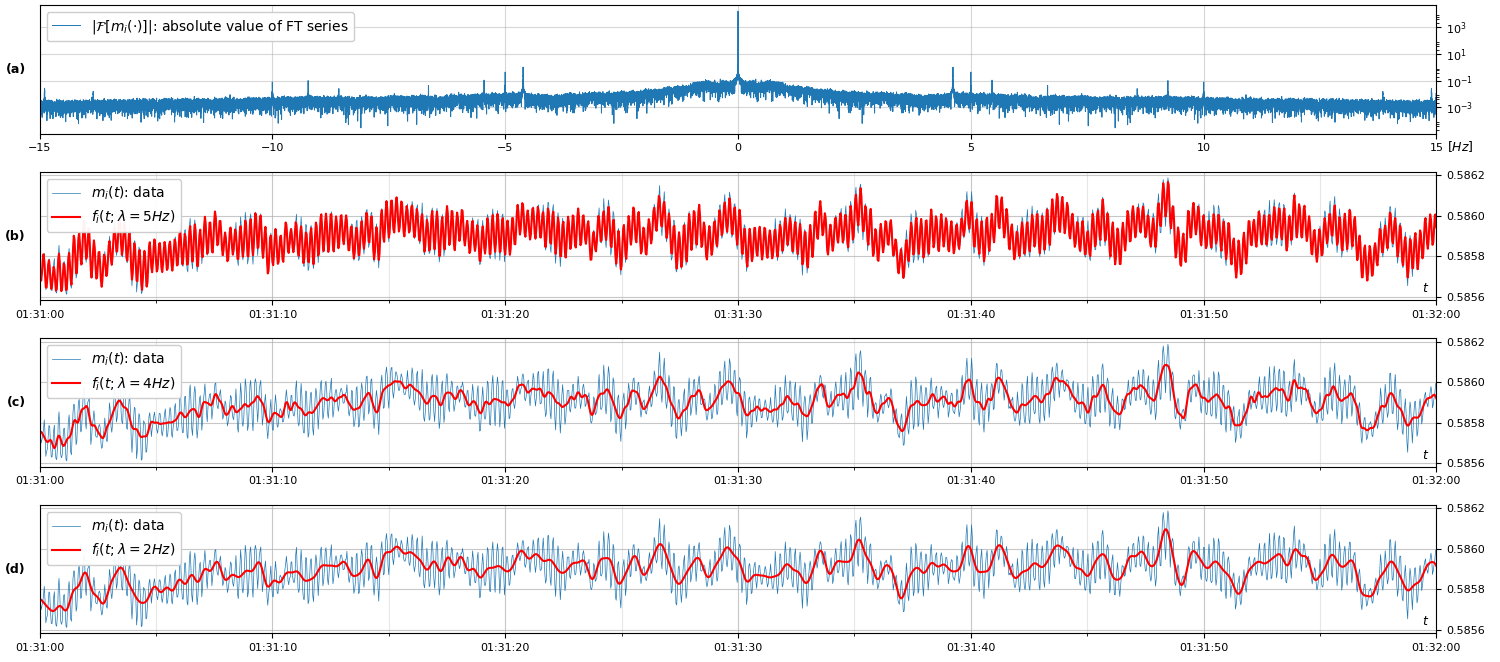}
  \caption{{\bf (a)} Absolute value (in log-scale) of Fourier transform of the voltage magnitude time series for a particular bus. Raw data and filtered series at {\bf (b)} $\lambda=5$Hz, {\bf (c)} $\lambda=4$Hz, and {\bf (d)} $\lambda=2$Hz. 
  \label{fig:ft-filter}}
\end{figure*}

Fig.~\ref{fig:ft-filter}a shows the amplitude (in absolute value) of the Fourier transform of the voltage of a bus for an interval of 15 minutes. 
Fig.~\ref{fig:ft-filter}b-d show the filtered series when frequencies larger than 5Hz, 4Hz, and 2Hz have been suppressed, respectively.

Analogously to equations~\eqref{eq:mu-s}-\eqref{eq:hat-m-s}, we can compute (averaging over a sliding time horizon) the zero-mean data vector and the re-scaled zero-mean data vector for $f_i(\cdot;\lambda)$ instead of $m_i(\cdot)$, we will call them $\bar{f}_i^{(s)}(\cdot;\lambda;S)$ and $\hat{f}_i^{(s)}(\cdot;\lambda;S)$, respectively. 



\begin{figure}
\centering  
  \includegraphics[width=.47\textwidth]{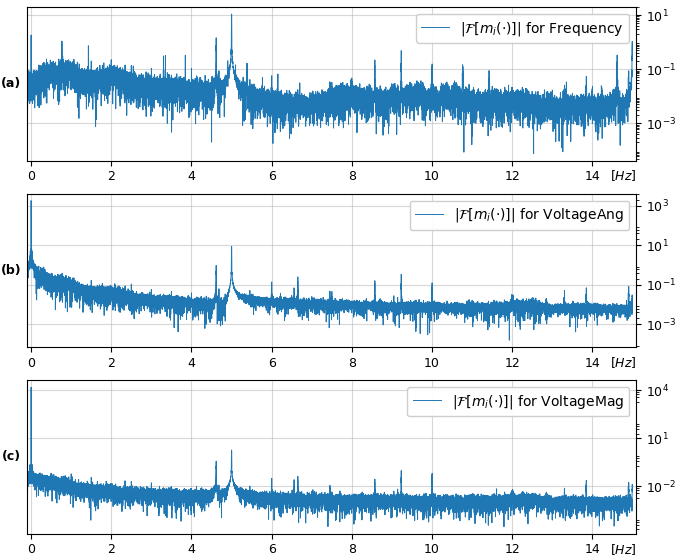}
  \caption{Fourier Transform of raw data for selected bus and selected period.} 
  \label{fig:fourier1}
\end{figure}

\begin{figure}
\centering  
  \includegraphics[width=.47\textwidth]{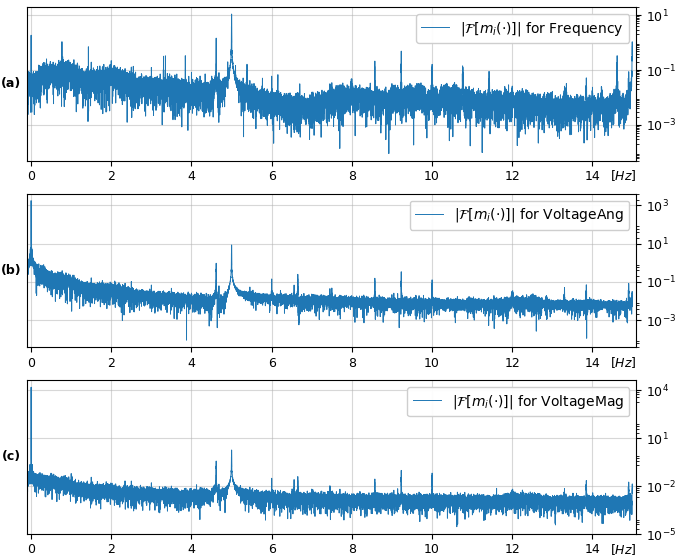}
  \caption{Fourier Transform of raw data for selected bus and selected period.} 
  \label{fig:fourier2}
\end{figure}

\begin{figure}
\centering  
  \includegraphics[width=.47\textwidth]{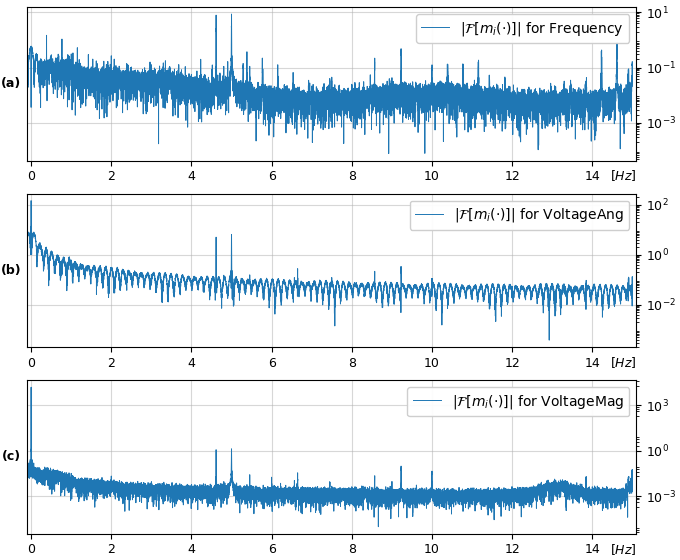}
  \caption{Fourier Transform of raw data for selected bus and selected period.} 
  \label{fig:fourier3}
\end{figure}

\begin{figure}
\centering  
  \includegraphics[width=.47\textwidth]{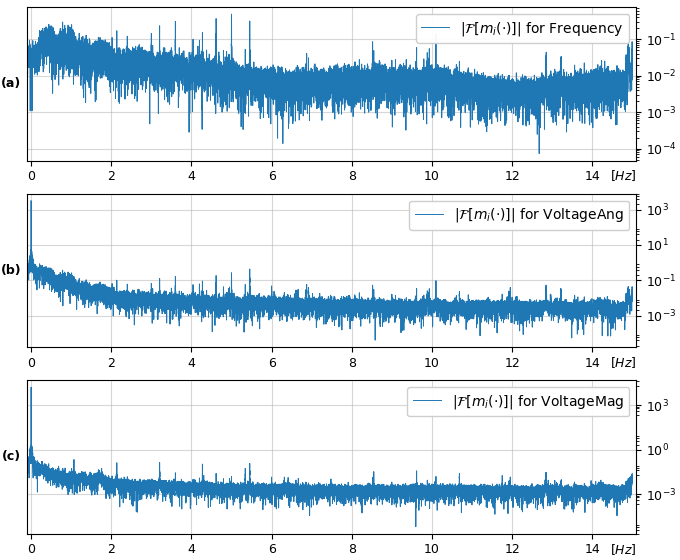}
  \caption{Fourier Transform of raw data for selected bus and selected period.} 
  \label{fig:fourier4}
\end{figure}

\begin{figure}
\centering  
  \includegraphics[width=.47\textwidth]{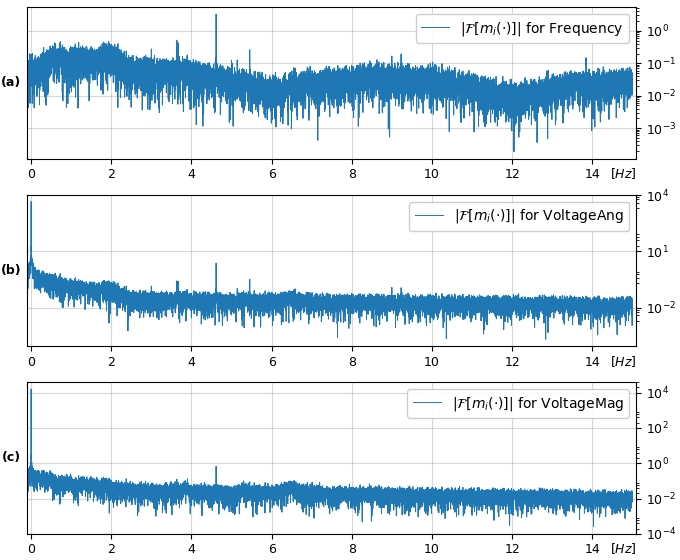}
  \caption{Fourier Transform of raw data for selected bus and selected period.} 
  \label{fig:fourier5}
\end{figure}

\begin{figure}
\centering  
  \includegraphics[width=.47\textwidth]{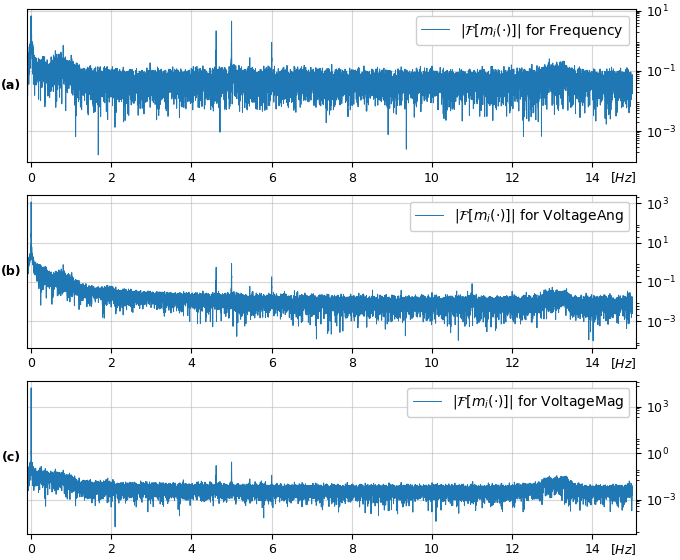}
  \caption{Fourier Transform of raw data for selected bus and selected period.} 
  \label{fig:fourier6}
\end{figure}

Figures~\ref{fig:fourier1}-\ref{fig:fourier6} plot the absolute value of the Fourier transform for positive frequencies (the negative side is symmetric since the original signal is real, not complex) and for three considered variables: (a) frequency, (b) voltage phase angle, and (c) voltage magnitude.

\section{Covariance Matrix \& PCA analysis}
\label{sec:PCA}

Let $T\leq Q$ be the number of measurements for a period of length greater than the parameter $S$. The $N\times N$ covariance (correlation) matrix of the signal, which is based on the last $T$ measurements, equals, at time $T\leq t\leq Q$:
\begin{align}
 \Sigma_0(t;T;m^*(\cdot))&=\Bigl[\ \frac{1}{T}\sum_{\tau=t-T+1}^t m_i^*(\tau) \overline{m_j^*(\tau)}\ \Big|\ \forall i,j\ \Bigr] \nonumber \\
&= \frac{1}{T}\sum_{\tau=t-T+1}^t m^*(\tau)m^*(\tau)^\H, \label{eq:Sigma}
\end{align}
where $m^*(\cdot)$ could be replaced by $\bar{m}^{(m)}(\cdot;\alpha)$,  $\hat{m}^{(m)}(\cdot;\alpha)$, $\bar{m}^{(s)}(\cdot;S)$,  $\hat{m}^{(s)}(\cdot;S)$, $\bar{f}^{(s)}(\cdot;\lambda;S)$, or $\hat{f}^{(s)}(\cdot;\lambda;S)$.

We aim to choose a (fixed) value $T$ that is sufficiently large so that the covariance is weakly dependent on $T$, but also not too large to keep memory as small as possible (with an eye on streaming applications).
Empirical experiments show that with $T$ corresponding to the number of measurements in three minutes we can obtain stable covariance matrices when $t$ varies.

In what follows we will drop the inputs $T$ and $m^*(\cdot)$ from $\Sigma_0(t;T;m^*(\cdot))$ since they become fixed in the following analysis. We perform an eigen-decomposition on the correlation matrices $\Sigma_0(t)$ and track the results as the function of $t$:
\begin{eqnarray}
\Sigma_0(t)=\sum_{k=1}^N \lambda_k(t) \xi_k(t)\xi_k(t)^\H, 
\label{eq:eigen-dec}
\end{eqnarray}
where $\xi_k(t)$ and $\lambda_k(t)$ are the the orthonormal eigenvectors and corresponding eigenvalues (in decreasing order) of $\Sigma_0(t)$ respectively, i.e. 
\begin{align}
 \forall k:\quad &\Sigma_0(t)\xi_k(t)=\lambda_k(t) \xi_k(t),\label{eq:eigen1}\\
 \forall k,l:\quad &\xi_k(t)^\H \xi_l(t)=\delta_{kl},\label{eq:eigen2}\\
& \lambda_1(t)\geq\lambda_2(t)\geq\cdots\geq\lambda_N(t)\geq0,\label{eq:eigen3}
\end{align}
where $\delta$ is the Kronecker delta function. Since $\Sigma_0(t)$ is a positive semidefinite matrix, \eqref{eq:eigen3} is justified.
In our tests, $\Sigma_0(t)$ is (numerically) rank-deficient, thus justifying the Principal Component Analysis (PCA) approach. PCA may be considered exact, if $K<N$ principal components are tracked or approximated. We approximate the sum in Eq.~\eqref{eq:eigen-dec} replacing,  $N=200$, by $K=10$. 

\begin{figure*}
  \includegraphics[width=\textwidth]{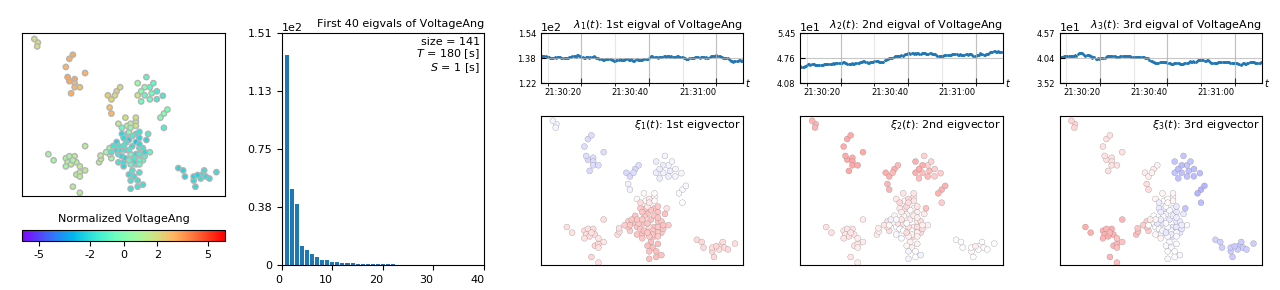}
  \caption{Movie screenshot of a selected quiet period at a specific moment $t$, showing the voltage phase angle PCA analysis. Here, $T=3$min. Columns from left to right: normalized phase angle (Eq.~\eqref{eq:hat-m-s}) at all buses, eigenvalues of $\Sigma_0(t)$, first, second, and third largest eigenvalues during the last minute (above) together with their corresponding normalized eigenvectors (below).}
  \label{fig:case1_phase}
\end{figure*}

The results of the PCA analysis are presented in Fig.~\ref{fig:case1_phase} which are screenshots of the movies available in~\cite{learn_pmu_web}. The screenshots and the movies show at time $t$ for each of the three variables (frequency, voltage angle, and voltage magnitude) five indicators:
\begin{itemize}
    \item normalized vector $m^*(t)$ of measurements: for each sensor, we plot at its geographical position the value of the normalized data using different colors for different values (column 1);
    \item first 40 eigenvalues of $\Sigma_0(t)$, in decreasing order (column 2);
    \item largest 3 eigenvalues $\lambda_1(\cdot),\lambda_2(\cdot),\lambda_3(\cdot)$ for the last minute before $t$ (columns 3 to 5, upper section);
    \item corresponding eigenvectors $\xi_1(t),\xi_2(t),\xi_3(t)$ at time $t$, component values of the vector are plotted geographically, in blue values close to $-1$ and in red values close to $1$ (columns 3 to 5, bottom section);
    \item auto-correlation function for a selected bus (column 6 in the movie---omitted in Fig.~\ref{fig:case1_phase}), this concept and figures are introduced in Section~\ref{sec:auto-correlations}.
\end{itemize}

For the numerical experiments we used traditional methods to compute the eigen-decomposition of $\Sigma_0(t)$, typically taking $O(N^3)$ time and using $O(NT)$ space. However, for streaming data that falls into a subspace of the original space (like in our case), we can consider lighter and faster algorithms to compute PCA, see \cite{17SYB}.

\begin{figure*}
  \includegraphics[width=\textwidth]{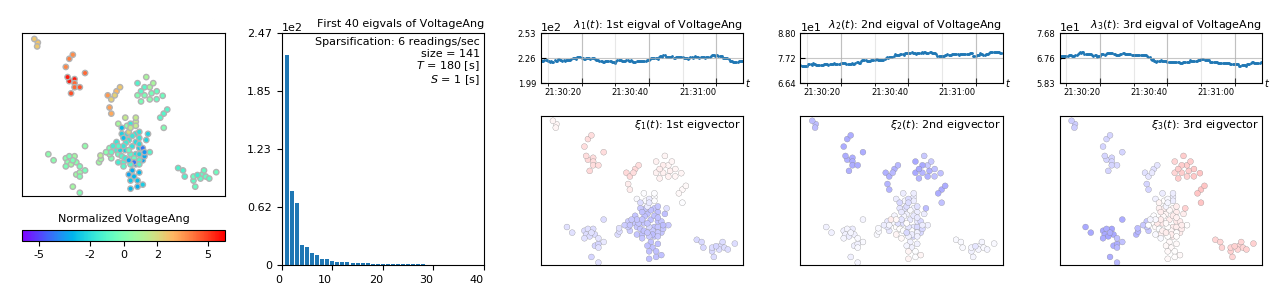}
  \caption{Movie screenshot of the same period and same time as Fig.~\ref{fig:case1_phase}. Here, the PCA analysis was made using only 6 readings per second.}
  \label{fig:case1_phase_sparse}
\end{figure*}

We also studied how the PCA changes if the data that its received is sparse, that is, from all the measurements that are made by the PMUs just a fraction of them are used in the analysis. 
As we mentioned above, PMUs report data 30 times per second, we performed PCA analyses with a fifth of that information, i.e. considering only 6 measurements per second and dropping the rest, meaning that each reading is equidistant in time by 0.167 [s].  Fig.~\ref{fig:case1_phase_sparse} shows the PCA analysis of the same period as Fig.~\ref{fig:case1_phase} for sparse measurements.
We observe that the same spectrum of eigenvalues is obtained, though the later analysis has higher eigenvalues. Moreover, the 3 eigenvalue functions that are shown have similar shape in Figs.~\ref{fig:case1_phase} and~\ref{fig:case1_phase_sparse}, as well as the eigen-vectors (omitting the sign change, flipping red and blue colors).
With this experiment we confirm what it is reported in previous literature \cite{18LWC}.

\subsection{Singular Value Decomposition (SVD)}

Consider the $N\times T$ measurement matrix 
$$M(t)=[\ M_{i,\tau}=m_i^*(\tau)\ |\ \forall i,\; \forall\tau\in\{t-T+1,\ldots,t\}\ ].$$ 
The SVD of $M(t)$ is
\begin{eqnarray}
M(t) = U(t) D(t) W(t)^\H, \label{eq:SVD}
\end{eqnarray}
where the ``spatial" matrix $U(t)$ is an $N\times N$ matrix the columns of which are orthogonal unit vectors of length $N$ which are called left singular vectors of $M(t)$, the ``temporal" matrix, $W(t)$, is $S\times S$ whose columns are right singular vectors of $M(t)$ and $D(t)$ is the $N\times S$ rectangular diagonal matrix of positive numbers. The covariance matrix, $\Sigma_0(t)$, is related to the measurement matrix, $M(t)$, according to 
\begin{eqnarray}
\Sigma_0(t)=M(t) M(t)^\H=U(t) D(t) D(t)^\H U(t)^\H, \label{eq:PCA-to-SVD}
\end{eqnarray}
where we took into account that $W(t)^\H W(t)=I$.
One observes that the covariance matrix does not depend on the temporal matrix $W(t)$.  To investigate time-related correlation effects, in the next section we also study  auto-correlations. 

\section{Accounting for Temporal Correlations: Auto-Correlation and Cross-Correlation Residue Maps}
\label{sec:temp-correlations}

Consider the delayed covariance matrix generalizing Eq.~\eqref{eq:Sigma}, 
$\forall \Delta\geq0,\ \forall t\geq T+\Delta:$
\begin{align}
\Sigma_\Delta(t;T;m^*(\cdot)) &= \Bigl[\ \frac{1}{T}\sum_{\tau=t-T+1}^t m^*_i(\tau) \overline{m^*_j(\tau-\Delta)}\ \Big|\ \forall i,j\ \Bigr] \nonumber\\
&= \frac{1}{T} \sum_{\tau=t-T+1}^t m^*(\tau) m^*(\tau-\Delta)^\H. \label{Sigma_tau}
\end{align}
We are interested to study how $\Sigma_\Delta(t;T;m^*(\cdot))$ changes with $\Delta$ increase and then track evolution with $t$. 

However, evolution  of the spectrum (in the two-dimensional $(\Delta,t)$ space) is challenging.  Instead, we study two surrogate objects, introduced in the following two subsections, which are easier to visualize.
Again, since $T$ and $m^*(\cdot)$ are fixed, we will omit them as input of the $\Sigma_\Delta$ function.

\subsection{Auto-Correlation Functions} 
\label{sec:auto-correlations}
The normalized PMU's auto-correlation functions at different nodes are defined as follows:
    \begin{eqnarray}    
    \forall i: \qquad
    {\cal A}_i(\Delta;t)=\frac{[\Sigma_\Delta(t)]_{ii}}{[\Sigma_0(t)]_{ii}}.
    \label{eq:A_i}
    \end{eqnarray}
This object is of interest because of the following two reasons
\begin{itemize}
\item Dependence of the auto-correlation function on $\Delta$ indicates whether fluctuations around the mean at a particular node decay or not with time.  Stated differently this analysis tests if there are significant memory effects or if memory is lost. 
\item It accounts for the part of the measurement matrix which is ignored in the PCA analysis, as discussed above --that is, it accounts for the temporal matrix, $W$.
\end{itemize}

\begin{figure*}
  \includegraphics[width=\textwidth]{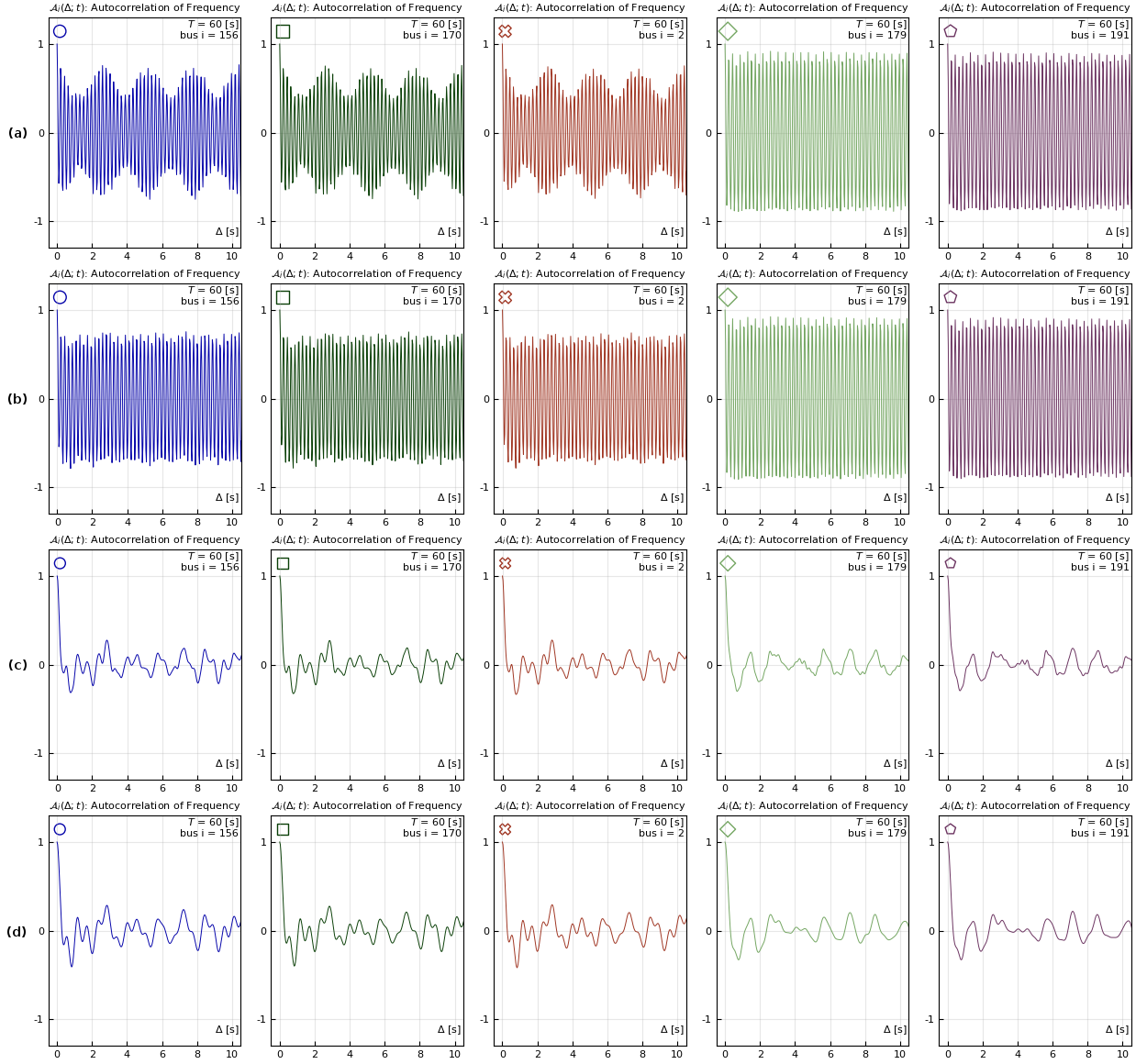}
  \caption{Movie screenshot of the auto-correlation functions for frequency at selected buses at a specific moment $t$. Here, $T=60$sec, and $S=1$sec.
  Different rows show the result using different time series: {\bf (a)} $\hat{m}^{(s)}(\cdot;S)$,  {\bf (b)} $\hat{f}^{(s)}(\cdot;\lambda=5Hz;S)$,  {\bf (c)} $\hat{f}^{(s)}(\cdot;\lambda=4Hz;S)$,  and {\bf (d)} $\hat{f}^{(s)}(\cdot;\lambda=2Hz;S)$.
  }
  \label{fig:autocorr_freq_fns}
\end{figure*}

We show dependence of the auto-correlations on time,  i.e. dependence of ${\cal A}_i(\Delta;t)$ on $t$. The movies are available in~\cite{learn_pmu_web} and example snapshots (of frequency) are given in row (a) of Fig.~\ref{fig:autocorr_freq_fns}, which shows five auto-correlation functions ${\cal A}_i(\cdot;t)$ for 5 selected buses at a specific time $t$. The buses that are shown were chosen between the ones that have larger auto-correlation amplitude. 

We also apply the auto-correlation analysis to the filtered signal.  The results of filtering the row signal, correspondent to the  Fig.~\ref{fig:autocorr_freq_fns}a, at 5Hz, 4Hz and 2Hz (cutting off higher harmonics) are shown in Fig.~\ref{fig:autocorr_freq_fns}b-d. We observe that if the signal is filtered at a sufficiently low frequency (4Hz) in this case a significant level of oscillations, seen otherwise in Fig.~\ref{fig:autocorr_freq_fns}a-b, disappear.

In order to measure this amplitude and formalize the concept of the \emph{residue} one studies 
\begin{eqnarray} \forall i: \qquad
\rho_i([\Delta_{min},\Delta_{max}];t)=\frac{\sum_{\Delta=\Delta_{min}}^{\Delta_{max}-1}|{\cal A}_i(\Delta;t)|}{\Delta_{\max}-\Delta_{min}},
\label{eq:A_i-res}
\end{eqnarray}
where $\Delta_{min}$ was set to $1s$ (in order to ignore the firsts values of ${\cal A}_i(\cdot;t)$) and $\Delta_{max}$ was set to $1min=60s$ in our tests. (We have verified empirically that with this value of $\Delta_{max}$ the results are robust with $O(1)$ changes in $\Delta_{max}$.)

\begin{figure*}
\centering  
  \includegraphics[width=\textwidth]{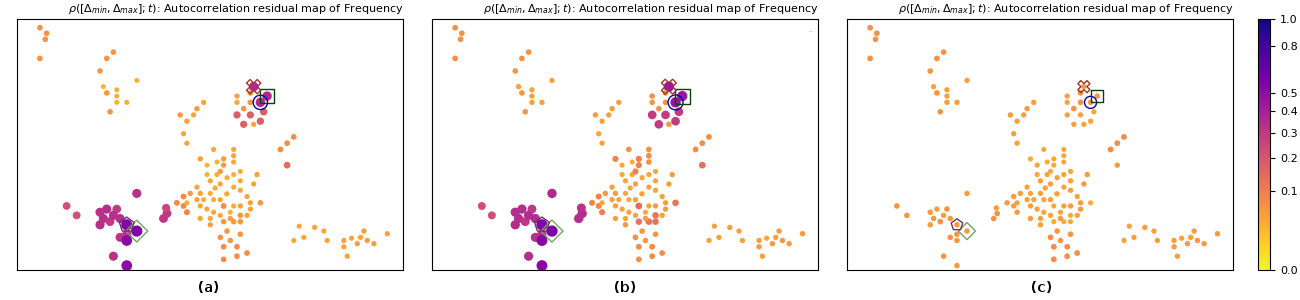}
  \caption{Movie screenshot of the auto-correlation map for frequency of Fig.~\ref{fig:autocorr_freq_fns}a-c, that is, using $\hat{m}^{(s)}(\cdot;S)$, $\hat{f}^{(s)}(\cdot;\lambda=5Hz;S)$, and $\hat{f}^{(s)}(\cdot;\lambda=4Hz;S)$, respectively. Here, $T=60$sec, $S=1$sec, $\Delta_{min}=1$sec, and $\Delta_{max}=60$sec. Color and size of the marks show the residue for each bus. Geometric figures (circle, square, cross, diamond and pentagon) correspond to the position of the buses whose auto-correlation function are shown in Fig.~\ref{fig:autocorr_freq_fns}.  }
  \label{fig:autocorr_freq_map}
\end{figure*}

Fig.~\ref{fig:autocorr_freq_map} geographically shows the residue for each node when different normalized time series are considered.
As seen in the movies of the residue maps shown in Fig.~\ref{fig:autocorr_freq_map}, residue values at a number of special nodes do not decay with time. Moreover we observe that nodes with a significant residue cluster, specifically, there are two sets of nodes that repeat their high values across different days, months and time of the day. Those groups are also shown in Fig.~\ref{fig:autocorr_freq_map}a: in the bottom left part of the map and the upper right zone (closer to the center); they might show up together or by separate.  

Observation that sustainable correlations stay for sufficiently long period of time suggest to analyze spatio-temporal features of the sustainable correlations via the Cross-Correlation Residue (CCR) described in the next Subsection. 

\subsection{Cross-Correlation Residue (CCR)}

\begin{figure*}
\centering  
  \includegraphics[width=.9\textwidth]{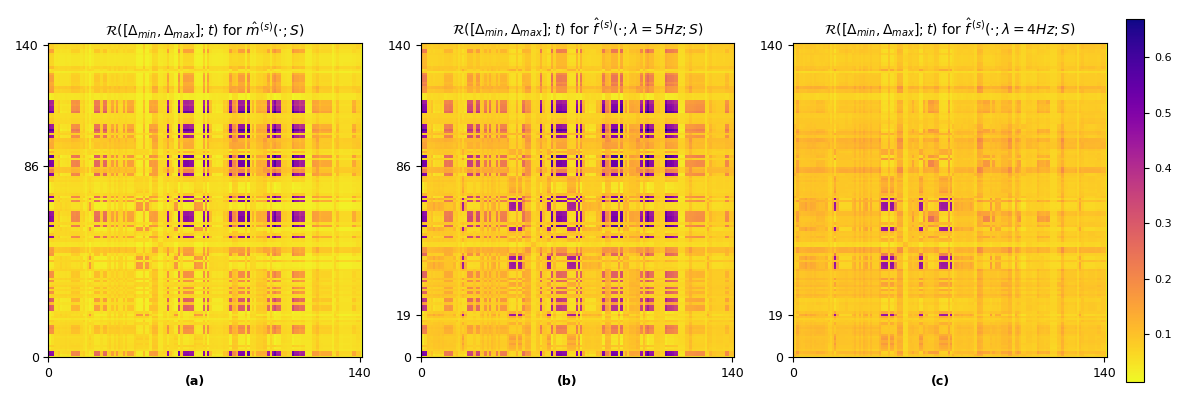}
  \caption{Cross-correlation matrices of 141 buses using {\bf (a)} $\hat{m}^{(s)}(\cdot;S)$, {\bf (b)} $\hat{f}^{(s)}(\cdot;\lambda=5Hz;S)$, and {\bf (c)} $\hat{f}^{(s)}(\cdot;\lambda=4Hz,S)$. Here,  $T=60$sec, $S=1$sec, $\Delta_{min}=1$sec, and $\Delta_{max}=60$sec.}
  \label{fig:residue_freq_matrix}
\end{figure*}

The cross-correlation version of Eqs.~\eqref{eq:A_i}-\eqref{eq:A_i-res} is
\begin{align}
\forall i,j: \qquad \quad
{\cal B}_{ij}(\Delta;t) &=[\Sigma_\Delta(t)]_{ij},     \label{eq:B_ij}\\
{\cal R}_{ij}([\Delta_{min},\Delta_{max}];t)
 &= \frac{\sum_{\Delta=\Delta_{min}}^{\Delta_{max}-1} |{\cal B}_{ij}(\Delta;t)|}{\Delta_{max}-\Delta_{min}},
\label{eq:B_ij-res}
\end{align}
where we drop normalization to avoid singularities associated with signals at different nodes which are not correlated. To visualize the CCR \eqref{eq:B_ij-res}, we plot the matrix ${\cal R}([\Delta_{min},\Delta_{max}];t) = \left[{\cal R}_{ij}([\Delta_{min},\Delta_{max}];t)|\forall i,j\right]$ showing in darker colors the higher values, see Fig.~\ref{fig:residue_freq_matrix}.

The three plots in Fig.~\ref{fig:residue_freq_matrix} show the CCR matrix at the same time using different time series, and depending on this, different patterns appear. In Fig.~\ref{fig:residue_freq_matrix}a 27 buses have in its corresponding row some component with a value higher than 0.4, whereas in Fig.~\ref{fig:residue_freq_matrix}c only 13 buses and different from the ones in Fig.~\ref{fig:residue_freq_matrix}a. Fig.~\ref{fig:residue_freq_matrix}b shows 40 buses with values higher than 0.4, the same as Fig.~\ref{fig:residue_freq_matrix}a and Fig.~\ref{fig:residue_freq_matrix}c together.

\begin{figure}
\centering  
  \includegraphics[width=.48\textwidth]{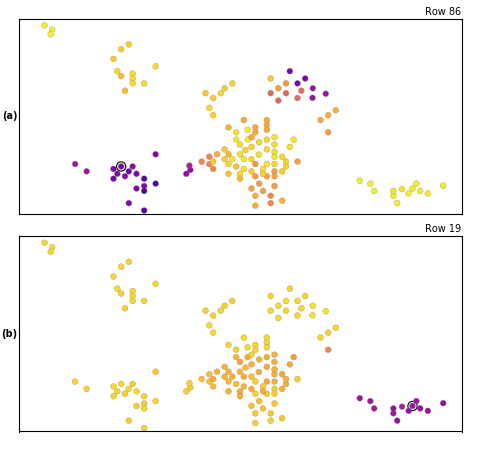}
  \caption{Geographical map with values of {\bf (a)} row 86 of Fig.~\ref{fig:residue_freq_matrix}a and {\bf (b)} row 19 of Fig.~\ref{fig:residue_freq_matrix}c. Buses corresponding to each row have an extra outer-circle.} 
  \label{fig:residue_freq_map}
\end{figure}

Fig.~\ref{fig:residue_freq_map} shows geographically the value that each bus has in row 86 of Fig.~\ref{fig:residue_freq_matrix}a and row 19 of Fig.~\ref{fig:residue_freq_matrix}c. We can see from these plots that high cross-correlation between buses is strongly connected with geographical closeness.
We also note that components that have high auto-correlation in Fig.~\ref{fig:autocorr_freq_map} have also high cross-correlation between themselves.

\section{Conclusions: What does online PCA and Auto-Correlations show?}
\label{sec:conclusions}

In this Section we summarize ``detective" methodology developed in the manuscript and the results it has helped to discover analyzing ``quiet periods" in the OGE data.  We will start with the latter (results) and then present a brief summary for the toolbox of the ``detective" methods.

\underline{\it Highlights of the Results:} 
\begin{itemize}
    \item Averaged over time PMU signal shows interesting spatial correlations. Correlations are different for different objects of interest (frequency, phase and voltage) and also different for different quiet periods. Matrix of correlations is sparse, also revealing that number  of the high-intensity contributions to the correlations is small. (Note that the statement of sparseness is consistent with previous studies of the measurement matrix, see e.g. \cite{18LWC}.) Each of the contributions characterizes a mode localized on a relatively few nodes (PMU positions) within the system. Principal modes, computed over the quiet periods, are almost frozen in time, however responding fast to any significant perturbation,  thus suggesting them as efficient features/indicators for changes. 
    
    \item Fourier-analysis of the signal reveals interesting spatial patterns. Extracting modes in the 4-6Hz range one observes significant contributions from only few PMU nodes.  The spatial patterns are different for different characteristics (frequency, phase, voltage amplitude) and they are also different for different quiet periods. Nodes showing large 4-6Hz contributions were identified as aggregated loads, mid-size generators and large wind-farms.  
    See Figures~\ref{fig:fourier1}-\ref{fig:fourier6} for Fourier analysis on a selection of nodes.
    
    \item In general the signals are long-correlated in time. However the memory effects becomes significantly less prominent when Fourier filtering at 4Hz, cutting higher frequencies, is applied. Nodes with significant auto-correlations (memory) are relatively few and the respective spatial pattern is adiabatic/frozen (changing slowly during the quiet periods). Like other adiabatic patterns mentioned above,  the pattern changes from one quiet period to another and we also observed different patterns for different characteristics (frequency, phase, voltage).  
    
    \item Analyzing time-delayed cross-correlations between different nodes we observe that, like in the case of the auto-correlations, correlations between some nodes have long memory. Nodes which mutual inference shows a long memory form a sparse pattern. These patterns, like others described above are adiabatic and evolving from characteristic to characteristic and from one quiet period to another quiet period.

\end{itemize}

\underline{\it\ Summary of Methods:} 
\begin{itemize}
    \item {\bf Averaging} -- computation of mean and variance taking into account recent history, exponential moving average and averaging over sliding time horizon methods are used and compared; with these statistics we compute normalized time series.
    \item {\bf Fourier Analysis} -- computation of discrete-time Fourier transform over the samples of an entire quiet period obtaining a new representation of the data, in the frequency domain; we obtain filtered time series by suppressing high frequency components (in the frequency domain) and applying the inverse Fourier transform.
    \item {\bf PCA} -- study of the eigen-decomposition of the covariance (correlation) matrix defined by the normalized time series; we plot how the spectrum (of eigenvalues) changes over time, as well as leading eigenvalues and eigenvectors.
    \item {\bf Auto-Correlation Analysis} -- computation of auto-correlation functions that explain how correlated are intervals of measurements at a node with previous intervals, we define the residue at a node to be related with the amplitude of the auto-correlation function.
    \item {\bf Cross-Correlation Residue} -- same as later, but now the correlation is in between an interval of measurements at a node with previous history at a different node.
\end{itemize}

\section{Path Forward}
\label{sec:path-forward}

In this manuscript we have developed algorithms that use PMU data for real-time separation of quiet periods from anomalous behavior. We argue that  tracking  algorithms of the type described above, validated on historical data from system wide PMU measurements, will make it possible in the future to extract and interpret hidden equipment malfunction and other extraneous phenomena. For example,  we should be able to pin down sources of forced oscillations at different frequencies (e.g. in the range from 0.5Hz to 5Hz) to malfunction of transformers, generators or large wind farms in less than a minute after  hard-to-identify small signatures of a potentially big future problems appear. We anticipate that accurate identification of the hidden signatures of the malfunctions will also be possible through introduction of learning, based on physical models of the system, in particular of a generalized swing-equation type discussed in \cite{2017DTCS,2017TDLCS} and possibly their nonlinear versions. The basic tools of statistical analysis, we took advantage of in the manuscript,  will be empowered in the future through injection physical models of the underlying phenomena into the purely statistical description of this manuscript. We also anticipate that fusing parameterized physics-based models for elements of the system we have sufficient amount of information (e.g. generators, transformers, well understood and aggregated traditional loads) with Neural Networks, e.g. of the modern DL type~\cite{09Ben}-\cite{16GBC}, for elements and parts of the system which we know much less about (renewable sources, active loads, etc) will be most beneficial in practice.  

Our ``detective'' approach to analysis of the quiet periods, started in the paper, will become even more critical for the next step -- accurate detection, localization and classification of more sever disturbances/events almost instantaneously (within seconds) after their occurrence. Overall we envision devising modern, physics of power systems informed machine learning methodology based but also going beyond what is the state of the art in the physics-blind and application agnostic IT industry, such as deep learning solutions. To validate the new approaches data-partnership with multiple ISOs and regional grid owners will be imperative for future success.

\appendix

\section{Primer of Auto-Correlations}

In this Appendix we discuss meaning of different features of auto-correlations on specific dynamic, scalar, real processes. We discuss here continuous time processes (transition to discrete time is obvious) stochastic and deterministic processes.
In the following we use the following notations for averaging over time, $\frac{1}{T}\int_0^Tdt A(t)\xrightarrow[T\to\infty]{}\mathbb{E}[A(t)]$. 

\subsection{Stochastic Colored Gaussian Process}

The process is defined through the following stochastic differential (Langevien) equation
\begin{eqnarray}
\frac{d}{dt} x(t)=-\frac{x}{\tau} +\xi(t),
\label{eq:color}
\end{eqnarray}
where $\xi(t)$ is zero mean white Gaussian short ($\delta$-)correlated noise, $\mathbb{E}[\xi(t)]=0,\quad \mathbb{E}[\xi(t)\xi(t')]=D\delta(t-t')$. Then in the statistically stationary regime (initial conditions are forgotten and the process is running for a while) the auto-correlation function of $x(t)$ is
\begin{eqnarray}
\mathbb{E}[x(t)x(t')]=\frac{D}{\tau}\exp\left(-\frac{|t-t'|}{\tau}\right). \label{eq:auto-color}
\end{eqnarray}

\subsection{Deterministic Periodic Process} 

Consider 
\begin{eqnarray}
x(t)=\sum_{n=1}^N a_n\cos\left(\frac{t}{\tau_n}+\theta_n\right).
\label{eq:cos}
\end{eqnarray}
Respective auto-correlation function is 
\begin{eqnarray}
\frac{1}{T}\int_0^T dt' x(t')x(t'+t)\xrightarrow[T\to\infty]{}\sum_{n=1}^N \frac{a^2_n}{2}\cos \left(\frac{t}{\tau_n }\right).
\label{eq:auto-cos}
\end{eqnarray}

\section*{Acknowledgments} The work at LANL was carried out under the auspices of the National Nuclear Security Administration of the U.S. Department of Energy under Contract No. DE-AC52-06NA25396. The work was partially supported by DOE/OE/GMLC and LANL/LDRD/CNLS projects.

\bibliographystyle{IEEEtran}
\bibliography{LearnPMU}

\begin{thebibliography}{10}
\providecommand{\url}[1]{#1}
\csname url@samestyle\endcsname
\providecommand{\newblock}{\relax}
\providecommand{\bibinfo}[2]{#2}
\providecommand{\BIBentrySTDinterwordspacing}{\spaceskip=0pt\relax}
\providecommand{\BIBentryALTinterwordstretchfactor}{4}
\providecommand{\BIBentryALTinterwordspacing}{\spaceskip=\fontdimen2\font plus
\BIBentryALTinterwordstretchfactor\fontdimen3\font minus
  \fontdimen4\font\relax}
\providecommand{\BIBforeignlanguage}[2]{{%
\expandafter\ifx\csname l@#1\endcsname\relax
\typeout{** WARNING: IEEEtran.bst: No hyphenation pattern has been}%
\typeout{** loaded for the language `#1'. Using the pattern for}%
\typeout{** the default language instead.}%
\else
\language=\csname l@#1\endcsname
\fi
#2}}
\providecommand{\BIBdecl}{\relax}
\BIBdecl

\bibitem{70SW}
F.~Schweppe and J.~Wildes, ``{Power System Static-State Estimation, Part I:
  Exact Model},'' \emph{IEEE Trans. on Power Apparatus and Systems}, vol.
  PAS-89, no.~1, pp. 120--125, Jan 1970.

\bibitem{70SR}
F.~Schweppe and D.~Rom, ``{Power System Static-State Estimation, Part II:
  Approximate Model},'' \emph{IEEE Trans. on Power Apparatus and Systems}, vol.
  PAS-89, no.~1, pp. 125--130, Jan 1970.

\bibitem{70Sch}
F.~Schweppe, ``{Power System Static-State Estimation, Part III:
  Implementation},'' \emph{IEEE Trans. on Power Apparatus and Systems}, vol.
  PAS-89, no.~1, pp. 130--135, Jan 1970.

\bibitem{1990Hauer}
J.~Hauer, C.~Demeure, and L.~Scharf, ``{Initial results in Prony analysis of
  power system response signals},'' \emph{IEEE Trans. on Power Systems},
  vol.~5, no.~1, pp. 80--89, Feb 1990.

\bibitem{2007Mani}
G.~Liu, J.~Quintero, and V.~Venkatasubramanian, ``Oscillation monitoring system
  based on wide area synchrophasors in power systems,'' in \emph{2007 iREP
  Symposium - Bulk Power System Dynamics and Control - VII. Revitalizing
  Operational Reliability}, Aug 2007, pp. 1--13.

\bibitem{12Tru}
D.~Trudnowski, ``{Properties of the Dominant Inter-Area Modes in the WECC
  Interconnect},''
  \url{https://www.wecc.biz/Reliability/WECCmodesPaper130113Trudnowski.pdf},
  2012, accessed: 2018-09-11.

\bibitem{13KXG}
M.~Kezunovic, L.~Xie, and S.~Grijalva, ``The role of big data in improving
  power system operation and protection,'' in \emph{2013 IREP Symposium Bulk
  Power System Dynamics and Control - IX Optimization, Security and Control of
  the Emerging Power Grid}, Aug 2013, pp. 1--9.

\bibitem{18AZ}
R.~Arghandeh and Y.~Zhou, \emph{{Big Data Application in Power Systems}}.\hskip
  1em plus 0.5em minus 0.4em\relax Elsevier, 2018.

\bibitem{DOE_call}
``Department of energy: Big data analysis of synchrophasor data,''
  \url{https://grantbulletin.research.uiowa.edu/netl-big-data-analysis-synchrophasor-data},
  2018.

\bibitem{NERC_reliability}
``Reliability {G}uideline {F}orced {O}scillation {M}onitoring \&
  {M}itigation,''
  \url{https://www.nerc.com/pa/RAPA/rg/ReliabilityGuidelines/Reliability_Guideline_-_Forced_Oscillations_-_2017.pdf
  }, 2017, accessed: 2018-09-11.

\bibitem{PNNL_oscil}
``{Power System Oscillatory Behaviors: Sources, Characteristics, \&
  Analyses},''
  \url{https://www.pnnl.gov/main/publications/external/technical_reports/PNNL-26375.pdf},
  2017, accessed: 2018-09-11.

\bibitem{WECC-JSI}
``{Report of {WECC} Joint Synchronized Information Subcommittee on Modes of
  Inter-Area Power Oscillations in Western Interconnection},''
  \url{https://www.wecc.biz/Reliability/WECC%20JSIS%20Modes%20of%20Inter-Area%20Oscillations-2013-12-REV1.1.pdf},
  2017, accessed: 2018-09-11.

\bibitem{12DKM}
N.~Dahal, R.~King, and V.~Madani, ``Online dimension reduction of synchrophasor
  data,'' \emph{Transmission and Distribution Conference and Exposition
  (T\&D)}, pp. 1--7, May 2012.

\bibitem{14XCK}
L.~Xie, Y.~Chen, and P.~Kumar, ``{Dimensionality Reduction of Synchrophasor
  Data for Early Event Detection: Linearized Analysis},'' \emph{IEEE Trans. on
  Power Systems}, vol.~29, no.~6, pp. 2784--2794, 2014.

\bibitem{16_RPI_missing_data}
P.~Gao, M.~Wang, S.~Ghiocel, J.~Chow, B.~Fardanesh, and G.~Stefopoulos,
  ``{Missing Data Recovery by Exploiting Low-Dimensionality in Power System
  Synchrophasor Measurements},'' \emph{IEEE Trans. on Power Systems}, vol.~31,
  no.~2, pp. 1006--1013, March 2016.

\bibitem{16_RPI_cyber}
P.~Gao, M.~Wang, J.~Chow, S.~Ghiocel, B.~Fardanesh, G.~Stefopoulos, and
  M.~Razanousky, ``{Identification of Successive “Unobservable” Cyber Data
  Attacks in Power Systems Through Matrix Decomposition},'' \emph{IEEE Trans.
  on Signal Processing}, vol.~64, no.~21, pp. 5557--5570, Nov 2016.

\bibitem{18LWC}
W.~Li, M.~Wang, and J.~Chow, ``{Real-Time Event Identification Through
  Low-Dimensional Subspace Characterization of High-Dimensional Synchrophasor
  Data},'' \emph{IEEE Trans. on Power Systems}, vol.~33, no.~5, pp. 4937--4947,
  Sept 2018.

\bibitem{18RL}
S.~Roy and B.~Lesieutre, ``{Frequency Band Decomposition of a Dynamic
  Persistence Measure Using Ambient Synchrophasor Data},'' in \emph{2018 Power
  Systems Computation Conference (PSCC)}, June 2018, pp. 1--7.

\bibitem{2009Chang}
\BIBentryALTinterwordspacing
C.~Chang and Z.~Li, ``Recursive stochastic subspace identification for
  structural parameter estimation,'' vol. 7292, 2009, pp. 7292 -- 7292 -- 9.
  [Online]. Available: \url{https://doi.org/10.1117/12.815422}
\BIBentrySTDinterwordspacing

\bibitem{2016Mani}
S.~Sarmadi and V.~Venkatasubramanian, ``{Inter-Area Resonance in Power Systems
  From Forced Oscillations},'' \emph{IEEE Trans. on Power Systems}, vol.~31,
  no.~1, pp. 378--386, Jan 2016.

\bibitem{2017LVSDC}
A.~{Lokhov}, M.~{Vuffray}, D.~{Shemetov}, D.~{Deka}, and M.~{Chertkov},
  ``{{Online Learning of Power Transmission Dynamics}},'' \emph{PSCC 2018,
  arXiv:1710.10021}, 2017.

\bibitem{09Ben}
\BIBentryALTinterwordspacing
Y.~Bengio, ``{Learning Deep Architectures for AI},'' \emph{Found. Trends Mach.
  Learn.}, vol.~2, no.~1, pp. 1--127, Jan. 2009. [Online]. Available:
  \url{http://dx.doi.org/10.1561/2200000006}
\BIBentrySTDinterwordspacing

\bibitem{10ARK}
I.~Arel, D.~Rose, and T.~Karnowski, ``{Deep Machine Learning - A New Frontier
  in Artificial Intelligence Research [Research Frontier]},'' \emph{IEEE
  Computational Intelligence Mag.}, vol.~5, no.~4, pp. 13--18, Nov 2010.

\bibitem{15Sch}
\BIBentryALTinterwordspacing
J.~Schmidhuber, ``{Deep Learning in Neural Networks},'' \emph{Neural Netw.},
  vol.~61, no.~C, pp. 85--117, Jan. 2015. [Online]. Available:
  \url{http://dx.doi.org/10.1016/j.neunet.2014.09.003}
\BIBentrySTDinterwordspacing

\bibitem{15LBH}
\BIBentryALTinterwordspacing
Y.~LeCun, Y.~Bengio, and G.~Hinton, ``Deep learning,'' \emph{Nature}, vol. 521,
  no. 7553, pp. 436--444, May 2015. [Online]. Available:
  \url{http://dx.doi.org/10.1038/nature14539}
\BIBentrySTDinterwordspacing

\bibitem{16GBC}
I.~Goodfellow, Y.~Bengio, and A.~Courville, \emph{Deep Learning}.\hskip 1em
  plus 0.5em minus 0.4em\relax The MIT Press, 2016.

\bibitem{2018attack}
D.~{Bienstock} and M.~{Escobar}, ``{Stochastic Defense Against Complex Grid
  Attacks},'' \emph{arXiv:1807.06707}, 2018.

\bibitem{learn_pmu_web}
\BIBentryALTinterwordspacing
D.~Bienstock, M.~Chertkov, and M.~Escobar, ``{Learning from ISO-scale PMU data
  stream},'' 2018. [Online]. Available:
  \url{http://www.columbia.edu/~dano/research/pgrid/pmus/index.html}
\BIBentrySTDinterwordspacing

\bibitem{17SYB}
\BIBentryALTinterwordspacing
A.~Shukla, S.-Y. Yun, and D.~Bienstock, ``{Non-Stationary Streaming PCA},''
  \emph{NIPS Time-Series Workshop}, 2017. [Online]. Available:
  \url{https://sites.google.com/site/nipsts2017/accepted-papers}
\BIBentrySTDinterwordspacing

\bibitem{2017DTCS}
D.~{Deka}, S.~{Talukdar}, M.~{Chertkov}, and M.~{Salapaka}, ``{Topology
  Estimation in Bulk Power Grids: Guarantees on Exact Recovery}.''

\bibitem{2017TDLCS}
\BIBentryALTinterwordspacing
S.~Talukdar, D.~Deka, B.~Lundstrom, M.~Chertkov, and M.~V. Salapaka,
  ``={Learning Exact Topology of a Loopy Power Grid from Ambient Dynamics},''
  in \emph{Proceedings of the Eighth International Conference on Future Energy
  Systems}, ser. e-Energy '17.\hskip 1em plus 0.5em minus 0.4em\relax New York,
  NY, USA: ACM, 2017, pp. 222--227. [Online]. Available:
  \url{http://doi.acm.org/10.1145/3077839.3077851}
\BIBentrySTDinterwordspacing

\end{thebibliography}
\end{document}